\documentstyle[aps,12pt,epsf]{revtex}

\begin{document} 
\baselineskip 12pt
\draft
\title{Geometric Solutions for the Neutrino 
Oscillation Length Resonance}

\author{Jason Pruet and George M. Fuller}
\address{Department of Physics, University of California,
San Diego, La Jolla, California 92093-0319}

\maketitle
\vskip 1.0cm

\begin{abstract}
We give a geometric interpretation of the neutrino \lq\lq oscillation 
length resonance \rq\rq\ recently discovered by Petcov. We use this 
picture to identify two new solutions for oscillation length resonances 
in a 3-layer earth model.
\end{abstract}
\newpage
\bigskip

In a recent study of the expected day-night asymmetry in the observed
 solar neutrino
spectrum  Petcov pointed out that neutrinos propagating
 through 
regions of varying density may undergo what has been called 
an \lq\lq oscillation length resonance \rq\rq\
\cite{pet1,pet2}. 
This has interesting consequences for the day-night asymmetry. In 
particular it suggests a region of vacuum mixing angle $\theta _0$
and mass-squared difference $\delta m^2$ parameter space
where the asymmetry may be enhanced. Significantly, this region coincides with 
the values of $\sin^2(2\theta _0)$ and $\delta m^2$ that best explain the 
Kamiokande and SuperKamiokande data \cite{hata}.
 As Petcov describes, this suggests the
possibilty of detecting a signature of this resonance and hence a clear
 indication
of neutrino oscillations. 

In this brief paper we give a geometric 
interpretation of the resonance. These geometric considerations permit
 an easy derivation
of the resonance properties and allow us to find two new resonance solution 
regions. We use our results to explain some interesting
features of the calculated day-night asymmetry and also to
clarify some confusion which has arisen from the more difficult algebraic 
approach.

In the following we work in the flavor basis and consider mixing between two 
neutrino species. We choose  $|\nu _e\rangle ={1 \choose 0} $ and
  $|\nu _{\mu}\rangle ={0 \choose 1} $.
 The method we use is based on the well known observation that
in this basis the MSW Hamiltonian \cite{M,SW1}

\begin{equation}
H = {\Delta _0 \over 2} \left(\matrix{- \cos 2\theta _0+{\sqrt2 G_f N_e /     \Delta _0}&
\ \sin 2\theta _0\cr
                  \ \sin 2\theta _0
&\cos 2\theta _0-{\sqrt2 G_f N_e /     \Delta _0}\cr}\right)
\label{MSW} \end{equation}
can be written as $H={1\over 2} \Delta _n \bar\sigma \cdot \hat n$
and so has the same form as the generator of spatial rotations for 
two-spinors \cite{MS2}. In the above $\bar\sigma$
  are the Pauli matrices, $N_e$ is the electron number density,
$\Delta _n=2\sqrt{H_{11}^2+H_{12}^2}$ , $\hat n$ is a unit vector with the 
cartesian components 
$(1/\Delta _n)(\Delta _0\sin2\theta _0,0,- \Delta _0 \cos2\theta _0
+\sqrt2 G_f N_e)$ , $\Delta _0 = \delta m^2 / 2 E _{\nu}$, 
and  $\hat n \cdot \hat z = \cos2\theta _n$ where $\theta _n$ is the 
matter mixing angle. Let us represent a state $|\nu \rangle =
{\alpha \choose \beta} $ which satisfies $\bf{\bar{\sigma}} {\cdot} \hat p 
{\alpha \choose \beta} = {\alpha \choose \beta}$ for some unit 3-vector
$\hat p$ by $|\nu \rangle   = |\hat p\rangle $. The ambiguity in overall phase will not
be of concern here.

 The following points are easily verified.

i) A neutrino which begins in a state represented by $|\hat a\rangle $ evolves, 
after passing for a time $t$ through a region with constant electron number
 density $N_e$ into a state $|\hat b\rangle $ where $\hat b$ is obtained by rotating
$\hat a$ about $\hat n(N_e)$ through an angle $\phi = \Delta _n (N_e) t$.
 In an obvious notation this is represented by
$|\hat a \rangle  \mapsto U(t)|\hat a\rangle =|\hat b = R\{\hat n(N_e),\phi\} \hat a\rangle $, 
where R is a 3x3 rotation matrix.

ii) The transition probability is given by,
\begin{equation} |\langle \hat a | \hat b\rangle |^2 = (1/2)(1+\hat a \cdot \hat b).
 \label{asjd} \end{equation}

These points simply serve to verify that the evolution of the neutrino can be 
represented by a 3-vector precessing in a cone about $\hat n(N_e)$. The case
of a neutrino propagating through regions of different density $(N_{e} ^{(1)},
N_{e}^{(2)},...)$ can be represented by a vector rotating first about 
$\hat n(N_{e}^{(1)})$, then about $\hat n(N_{e}^{(2)})$ after the appropriate
 time,
 and so on.

We now use this picture to solve the oscillation length resonance problem
for neutrinos passing through the earth. In this scenario a neutrino passes 
through a low density mantle characterized by electron number density
$N_{e}^m$ and length $L _m$, a higher density core with 
$N_{e}^c$ and length $L_c$ and the lower density mantle again 
($N_{e}^m$,$L_m$). For the $\delta m^2$ and neutrino energies we 
consider, a neutrino which begins in the sun as a combination of the 
mass eigenstates $|\nu _1\rangle $ and $|\nu _2 \rangle $ arrives at the earth with these
components well separated spatially, owing to the difference in their 
propagation velocities \cite{MS3}. The relevant conversion probability in 
this case is 
$P _{\nu _1 \mapsto \nu _{\mu}}$. The aim is to find, for a given 
$N_{e}^c,N_{e}^m$ (or equivelantly matter mixing angles  $\theta _c$ 
and $\theta _m$) the 
$L_c$ and $L_m$ which maximize this transition probability. For illustration 
we first consider the easier problem of finding the resonances for the 
transition probabilty between flavor eigenstates, 
$P _{\nu _e \mapsto \nu _{\mu}}$. This corresponds to finding the 
$\phi _1$ and $\phi _2$ which maximize 
$-\hat z^T R(\hat  n _m,\phi _1)R(\hat n _c,\phi _2)
R(\hat n _m,\phi _1)\hat z$. Clearly, $R(\hat n _m,\phi _1) \hat z =  
R(\hat n _c,\phi ') \hat p$ for some $\phi '$ and $\hat p$
where $\hat p$ is in the x-z plane and satisfies $1\geq \hat p \cdot \hat z
\geq \cos(2 \theta _c - |2(\theta _c - 2 \theta _m)|)$. Therefore, the 
quantity we 
wish to maximize is 
$-\hat p ^T R(\hat n _c,\phi _2 +2 \phi ') \hat p$. This takes a maximum when
$\phi _2 + 2 \phi  ' = \pi$, and when $\hat p \cdot \hat n _c$ is as close to
zero as allowed by the above constraint on $\hat p$. This and some 
geometry completely solves the problem of finding the resonances for
 $P _{\nu _e \mapsto \nu _{\mu}}$. Similar considerations solve the 
$P _{\nu _1 \mapsto \nu _{\mu}}$ problem. 
 
The results divide into three
cases. In the following 
$P^{res} _{\nu _1 \mapsto \nu _\mu}$ is the resonance (maximum)
value of the transition
probability, obtained when $L_m = L^{res} _m$ and $L_c = L^{res} _c$.

	{\bf{I}})  For case {\bf{I}} we have, $  2 \theta _c
 - \theta _0 < \pi /2 $,

	\begin{equation} P^{res} _{\nu _1 \mapsto \nu _\mu} =
 \sin^2(2\theta _c - \theta _0), \label{joe} \end{equation}
\begin{equation}	  L^{res} _m ={c \over {\Delta _n ^{(mantle)}}}2n \pi,
\label{tom} \end{equation} 
      \begin{equation} L^{res} _c = {c \over \Delta _n ^{(core)}}(2m+1) \pi,  
\label{frank} \end{equation}
	for $m,n=0,1,2,...$. In this case there is no enhancement of the 
transition probability, as $P _{\nu _1 \mapsto \nu _\mu}$
 can take this value in 
the core alone.

	{\bf{II}}) For case {\bf{II}} we have 
$|2\theta _c - 4 \theta _m + \theta _0| <
\pi /2 < 2 \theta _c - \theta _0,$
	
	 \begin{equation} P^{res} _{\nu _1 \mapsto \nu _\mu} = 1,
 \label{a} \end{equation}
         \begin{equation}  L^{res} _m ={c \over {\Delta _n ^{(mantle)}}}
 (\rho + 2n\pi ), \label{b} \end{equation}
         \begin{equation} L^{res} _c = {c \over 
{\Delta _n ^{(core)}}}((2m+1)\pi -\gamma -\eta ), \label{g} \end{equation}
	 for $n,m=0,1,2,...$, and where 
\begin{equation} \cos\rho = {\cot(2(\theta _c - \theta _m))
(\cos 2\theta _m +\cos(2(\theta _m - \theta _0))) \over
(\sin 2\theta _m + \sin 2(\theta _m - \theta _0))}, \label{h} \end{equation} 
\begin{equation} \tan \gamma = {\sin \rho \over
(\cot 2\theta _m \sin 2(\theta _c - \theta _m) +  \cos \rho \cos 2 
(\theta _c - \theta _m))}, \label{k} \end{equation}
\begin{equation}  \tan \eta = {\sin \rho \over
(\sin 2(\theta _c - \theta _m) \cot 2(\theta _m -\theta _0) + 
 \cos \rho \cos 2(\theta _c - \theta _m))}, \label{q} \end{equation}
Note that if $\rho$ is a solution then $2\pi - \rho$ is also, with 
$\gamma \mapsto - \gamma , \eta \mapsto -\eta$.

{\bf{III}}) Petcov's original 
solution. $|2\theta _c - 4 \theta _m + \theta _0| > \pi/2$

\begin{equation} P^{res} _{\nu _1 \mapsto \nu _\mu} =  
\sin^2(2\theta _c + \theta _0 -4 \theta _m), \label{w} \end{equation}
\begin{equation}  L^{res} _m ={c \over {\Delta _n ^{(mantle)}}}(2n+1)\pi, 
\label{rw} \end{equation}
\begin{equation}  L^{res} _c = {c \over {\Delta _n ^{(core)}}}(2m+1)\pi,
\label{ssld} \end{equation}
for n,m=0,1,2,....   
This is the case discussed by Petcov in Ref. \cite{pet1}.

 Note that equations [12a,12b] of that paper, which give the conditions that
must be satisfied in order for the transition probability to take a local
maximum, differ from the conditions above because we are discussing global
maxima here. 
	
An interesting quantity is $P ^{res} - P^{one layer} _{max}$, the difference
between the parametric resonance transition probability and the maximum
that the transition probability can take in a single layer 
(defined to be max$\{ \sin ^2 [2 \theta _{inner} - \theta _0],
 \sin^2 [2 \theta _{outer}- \theta _0]\} $).
 This is plotted in Figure 1, where we 
have chosen the outer layers of the sandwich to have density ${\rho} = 
4.5\,{\rm g}{\rm cm}^{-3}$ and electron fraction $Y _e = .49$
and the inner layer to have $\rho=11.5{\rm g}{\rm cm}^{-3}$ and $Y_e =.46$. 
These are the average values
 given by the Stacey earth model and those
used by Petcov in \cite{pet1}. 

To illustrate these ideas we consider a particular case.
Suppose that $\theta _0 = .05$ and that for the neutrino energy we are 
interested in we satisfy
$\log [\delta m^2/2E_{\nu}({ev})] = -12.55$. Then by calculating the matter 
mixing angles we find that we are in Case {\bf {II}} above 
and that we can choose the lengths of the inner and
outer layers in such a way that the transition probability is unity. 
For this case $L ^{(outer)}=11629$km and 
 $L^{(inner)}=2080$
km. This is to be contrasted with the maximum that the transition probability could take in a single layer, which for
 this case is $\approx .53$.
Note that for a given $\theta _0$,  this type of 
 resonance only has an interesting effect 
in a narrow region about the MSW resonance energies in the mantle and core. 

The answer to the question of whether or not an oscillation length
 resonance actually 
 occurs in the 
earth depends on the actual 
core and mantle lengths and the extent to which this simple model of 
the earth mirrors reality. The question is most easily answered by simply 
examining $P ^{earth} - P ^{one \, layer}$. An examination of the 
dependence of $P ^{earth} - P ^{one \, layer}$ on $\theta _0$ and $E _{\nu}$
reveals that this quantity is 
positive in a small region. We conclude that there is destructive
 interference as often as constructive.
A different way of describing the phenomena in the earth then is to say not
 that it leads to 
an enhancement of the transition probabilty, but that the transition probability is more sensitive to 
the neutrino energy and phase of the neutrino state at the mantle/core boundary. 
Consequently, the transition probability for core-crossing neutrinos has a sharp energy dependence. It is 
characterized, for $2\theta _0 \sim .1$, by a peak centered at $E^{\rm {mantle}} _{\rm{res}} (1+.5)$, with a bump on 
each side. With a detector like SNO, with an expected energy resolution of $\sim 15\%$,
 this sharp structure may be useful for determining the mixing parameters. Of course, 
the clearer signature of MSW resonance in the mantle would first be observed.

In summary, we have shown how an analogy with the simple rotational
 geometry of  spin 1/2 systems can be used to picture and solve certain neutrino 
oscillation problems. We have illustrated the method
by finding the core and mantle lengths in a 3 layer earth model 
which maximize the $\nu _1 \mapsto \nu _{\mu}$
transition probability.  

We note that two papers by Chizhov and Petcov \cite{pet2,pet3} recently appeared which independently addressed 
some of the same issues investigated here. In contrast to our geometric
arguments they have algebraically derived the results for cases {\bf{I}}-{\bf{III}}.

{\bf{Acknowledgements.}} This work was supported in part by NSF Grant PHY98-00980 at UCSD. G.M.F.
acknowledges stimulating conversations with S. Petcov and A.B. Balantekin at the 1998 Neutrino Workshop 
at the Aspen Center for Physics.

\newpage
\noindent{\bf Figure Captions:}
\noindent
Figure 1.
Contour Plot of $P^{res} - P^{onelayer}$. The vertical axis is $2\theta _0$, the horizontal is $\log {{\delta m^2 / {2E}}}$ and the argument of the log
is in eV.

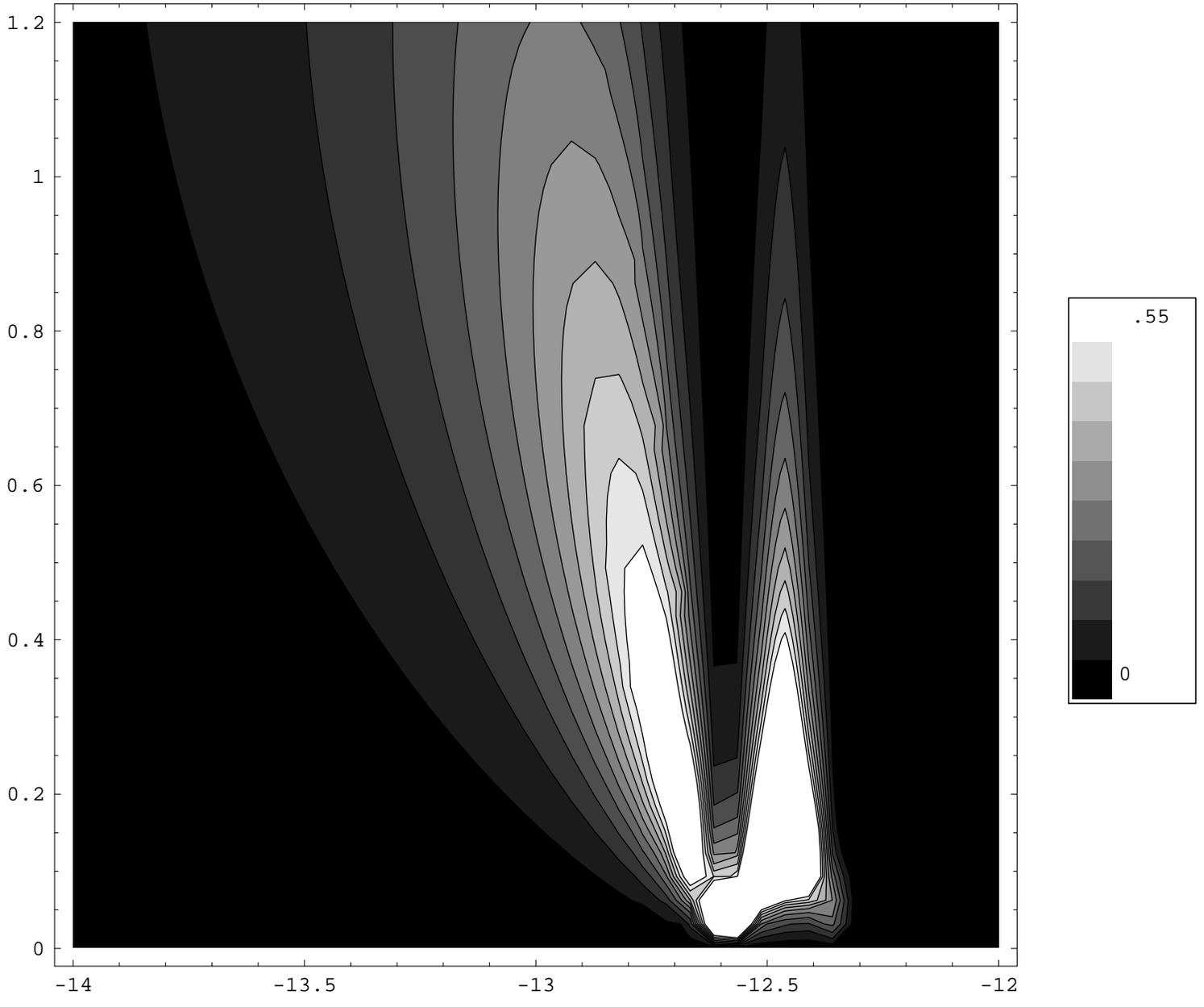
\end{document}